\documentclass[12pt,preprint]{aastex}
\def\lesssim{\mathrel{\hbox{\rlap{\hbox{\lower3pt\hbox{$\sim$}}}{\hbox{\raise2pt\hbox{$<$}}}}}}
\def\gtrsim{\mathrel{\hbox{\rlap{\hbox{\lower3pt\hbox{$\sim$}}}{\hbox{\raise2pt\hbox{$>$}}}}}}

\shorttitle{Furusho et al.}
\shortauthors{Temperature map of the Perseus Cluster}

\begin{document}

\title{Temperature Map of the Perseus Cluster of Galaxies Observed
with ASCA} \author{ T. Furusho \altaffilmark{1}, N. Y. Yamasaki
\altaffilmark{2}, T. Ohashi \altaffilmark{2}, R. Shibata
\altaffilmark{3}, and H. Ezawa \altaffilmark{4}
}

\altaffiltext{1}{Laboratory for High Energy Astrophysics, 
NASA/GSFC, Code 662, Greenbelt, MD 20771; furusho@olegacy.gsfc.nasa.gov}

\altaffiltext{2}{Department of Physics, Tokyo Metropolitan University, 
1-1 Minami-Ohsawa, Hachioji, Tokyo 192-0397, Japan}

\altaffiltext{3}{Institute of Space and Astronautical Science, 3-1-1,
  Yoshinodai, Sagamihara, Kanagawa 229-8510, Japan}

\altaffiltext{4}{Nobeyama Radio Observatory, National Astronomical
  Observatory, 462-2 Minamimaki, Minamisaku, Nagano 384-1305, Japan}

\begin{abstract}
We present two-dimensional temperature map of the Perseus cluster
based on multi-pointing observations with the ASCA GIS, covering a
region with a diameter of $\sim 2^\circ$. By correcting for the effect
of the X-ray telescope response, the temperatures were estimated from
hardness ratios and the complete temperature structure of the cluster
with a spatial resolution of about 100 kpc was obtained for the first
time.  There is an extended cool region with a diameter of $\sim 20'$
and $kT \sim 5$ keV at about $20'$ east from the cluster center. This
region also shows higher surface brightness and is surrounded by a
large ring-like hot region with $ kT \gtrsim 7$ keV, and likely to be
a remnant of a merger with a poor cluster.  Another extended cool
region is extending outward from the IC 310 subcluster. These features
and the presence of several other hot and cool blobs suggest that this
rich cluster has been formed as a result of a repetition of many
subcluster mergers.
\end{abstract}


\section{Introduction} 

Temperature maps of clusters of galaxies give almost direct knowledge
about the history of past subcluster mergers. Many clusters are now
known to exhibit significant temperature structures in the
intracluster medium (ICM) based mainly on ASCA observations, and these
results give us a view that clusters of galaxies are still evolving
(e.g.\ Markevitch et al.\ 1996, 1998). However, the spatial resolution
of ASCA is not capable of resolving $100h_{50}^{-1}$ kpc scale
structures for objects at $z \gtrsim 0.03$. On the other hand, nearby
clusters ($z< 0.03$) are very extended with $0.5 r_{\rm vir} >
1^\circ$, which requires us to perform multi-pointing
observations. This makes the analysis complicated because of the
energy-dependent point spread function (PSF) of the ASCA XRT, and only
recently several results have been coming out (Watanabe et al.\ 1999;
Kikuchi et al.\ 2000; Shibata et al.\ 2001; Furusho et al.\ 2001;
Watanabe et al.\ 2001).

This letter reports temperature structure of the Perseus cluster
(A426, $z=0.0183$) over a radius of $1.0^\circ$ ($\sim 2 h_{50}^{-1}$
Mpc).  This is the brightest cluster in the X-ray sky, and a very
suitable object for the study of various properties of the
intracluster medium (ICM)\@.  The PSPC image of the Perseus cluster
showed significant substructures in the east from the cluster center,
which has been interpreted as an evidence that the cluster is not in a
relaxed state but has undergone recent mergers (Schwarz et al.\ 1992;
Ettori et al.\ 1998). Following the initial ASCA study on the
temperature variation for the central pointing (Arnaud et al.\ 1994),
Ezawa et al.\ (2001) showed the large-scale metal abundance gradient
and fluctuation of temperature distribution in 4 sectors based on
mapping observations from ASCA\@.  They also reported an extended soft
component in the cluster center, characterized by a temperature $ kT
\sim 2$ keV\@. Recent Chandra image showed a complicated X-ray
morphology around the central galaxy NGC 1275 (Fabian et al.\ 2000).

In this letter, we report on the two-dimensional temperature map with
the resolution of $2.\!'5$ ($78 h_{50}^{-1}$ kpc) derived from
multi-pointing observations of the Perseus cluster with ASCA. We use
$H_0 = 50 h_{50}$ km s$^{-1}$ Mpc$^{-1}$ and $q_0 = 0.5$, which
indicates $1' \simeq 31$ kpc at the Perseus cluster.  The solar number
abundance of Fe relative to H is taken as $4.68 \times 10^{-5}$
(Anders and Grevesse\ 1989).

\section{Observation and Analysis}

ASCA observations of the Perseus cluster were performed in 4 separate
occasions between August 1993 and February 1997. The number of
pointings is 13 with an average exposure time of 10--20 ksec each, and
the total exposure time amounts to 172 ksec. The observed regions
cover a diameter of about $2^\circ$. In this letter, we only deal with
the GIS data because of its larger field of view than the SIS\@. The
data screening procedure was the same as in our previous analysis for
3 other clusters (Furusho et al.\ 2001).  The cosmic X-ray background
(CXB) was taken from the archival blank-sky data taken during
1993--1994 (Ikebe\ 1995).

The flow of the data analysis is almost the same as that described in
Furusho et al.\ (2001), i.e.\ we calculated hardness ratio in each
pixel of the merged image for the 13 pointings after the background
subtraction, and estimated temperature by comparing it with simulated
values for isothermal ICM\@. The ray-tracing simulation includes
effects from energy-dependent PSF and stray light (Tsusaka et al.\
1995). As for the input X-ray image of the cluster, we employed the
observed two-dimensional PSPC image in the energy range 0.5--2.0 keV
in the same way as in Furusho et al.\ (2001). The PSPC observation of
the Perseus cluster consists of 5 pointings, and the data were
analyzed with the Extended Source Analysis Software package (Snowden
et al.\ 1994). The input spectrum was assumed to be a
single-temperature MEKAL model with $kT = 6$ keV with the Galactic
absorption ($N_{\rm H} = 1.4\times10^{21}$ cm$^{-2}$), and the
significant metallicity gradient reported by Ezawa et al.\ (2001) was
included in the model.

The definition of the hardness ratio is ${\it HR} = H/S$, where $H$
and $S$ represent count rates in the energy range 2--10 keV and 0.7--2
keV, respectively.  The ratio of {\it HR}s between the data and the
isothermal simulation for each pixel gives a deviation factor of the
{\it HR} from the assumed temperature, along with its statistical
error. Note that the {\it HR} values for the isothermal cluster differ
from position to position because the PSF profile and the stray light
intensity have significant energy dependence. Based on the ${\it
HR}-kT$ relation, whose profile again depends on the position, we
estimated the temperature for each observed region.  The calculation
of the ${\it HR}-kT$ relation was based on the single temperature
model including the radial metallicity gradient. 

As for the pixel size, the hard and soft band images were first
constructed for $2.\!'5 \times 2.\!'5$ pixels. For each pixel, the
{\it HR} value used for the color-coded temperature map was calculated
for surrounding $4 \times 4$ cells ($10'\times 10'$). This running
average effectively suppresses the statistical fluctuation but reduces
the angular resolution at the same time. The statistical errors in the
$4 \times 4$ cells are all less than 10\%, and the systematic errors
due mainly to the fluctuation of the cosmic X-ray background and the
modelling of the XRT response (see Furusho et al.\ 2001) were
estimated to be about 20\% in the outer region and smaller in the
inner regions, respectively.

\section{Results}

Figure 1 shows the color-coded temperature map of the Perseus cluster,
compared with a residual PSPC surface brightness after subtracting a
smooth $\beta$ model from the observed image. The subtracted $\beta$
model parameters, derived for the smooth western sector, were taken
from Ettori et al.\ (1998). The temperatures were estimated assuming a
single temperature as mentioned earlier.  We first note that the
temperature structures in the map are generally consistent with the
previous coarse-resolution results by Ezawa et al.\ (2001). There is a
large cool region around $20'$ in the east from the cluster center,
which is clearly extended with a diameter of about $30'$ ($\sim
900h_{50}^{-1}$ kpc).  This cool region shows a remarkable
correspondence to the enhancement in the surface brightness as seen in
Figure\ 1.  The correspondence between the cool region and the excess
surface brightness suggests that this extended region is either
embedded in or lying foreground to the Perseus cluster. This
substructure in the surface brightness was already noted by Schwarz et
al.\ (1992) and Ettori et al.\ (1998), but this is the first case
which reveals that this feature accompanies a clear temperature drop.

The other interesting feature in the temperature map is the ring-like
hot region which almost encircles both the cool region and the cluster
center. The eastern ridge, at $r \sim 40'$ from the center running
north to south, seems particularly hot with $ kT \gtrsim 10$ keV in a
number of connected pixels. The rest of the ring-like feature
indicates approximately the same temperature with $kT \sim 7 $ keV\@.
There are other soft regions: one small region in the south-south-west
direction at $r \gtrsim 30'$, and an elongated region at the
north-east edge. Neither of the regions show structures in the surface
brightness distribution, suggesting that the low ICM temperatures in
these regions do not accompany significant density contrast.

The central region within a diameter of $10'$ shows a marked softening
of the spectrum, as studied by Ezawa et al.\ (2001).  They showed that
2 spectral components were needed to fit the spectra in the central
region within $r = 20'$, but the hotter component itself also
indicated a marked softening down to $kT = 5.7$ keV\@. We also carried
out spectral analysis assuming the 2-temperature model, and produced a
temperature map only for the hot component. As a result, the color
pattern turned out to be similar to Figure\ 1 in that the temperature
in the central region within $r = 7'$ became low with $kT< 5$
keV\@. See Ezawa et al.\ (2001) for more information about the
spectral softening in the central region.

Apart from the central concentration of the soft component, the two
bright point sources, IC 310 and 1RXS J031525.1+410620, in the
south-west region are clearly recognized in the temperature map. These
sources are characterized by low ($\sim 4$ keV) and high ($\sim 11$
keV) temperatures, respectively, showing that the temperature map can
effectively separate contributions from discrete sources.

One dimensional cutouts of the temperature distribution are shown in
Figure\ 2a along east-to-west and south-to-north paths running through
the center with a width of $10'$ as indicated in Figure\ 2b.  The
errors in the plot denote 90\% statistical errors for single
parameters. These plots clearly demonstrate that the observed
temperature variation is statistically significant. The extended cool
region is recognized as a flat region between $-20'$ and $-10'$ along
the path A in Figure 2a, which corresponds to the residual component
as shown by the contours in Figure 1. The very high temperature at the
eastern edge ($-40'$ in path A), the hot region in the south ($-30'$
in path B), and the northern cool region ($+35'$ in B) can also be
identified clearly.  These one dimensional plots suggest that a
significant change of the temperature occurs in a scale of $\sim 10'$
(300 kpc), even though the finer scale structure may be suppressed by
the bin width of $5'$.

To look into the spectral variation in correlation with the
temperature variation in some detail, we have accumulated the energy
spectra for selected regions and carried out spectral fits. We have
chosen 3 representative regions as shown in Figure\ 2b, which indicate
medium (\#1), low (\#2) and hot (\#3) temperatures, respectively. The
strong cool component around the central region causes the fit with
single temperature models unacceptable and complicates the temperature
estimation. To avoid this problem, we limited the energy range to
2--10 keV and carried out single temperature fits. Since the
temperature of the cool component is about 2.0 keV (Ezawa et al.\
2001), about 80\% of the photon flux in the cool component can be
suppressed. Free parameters were normalization, temperature and metal
abundance, and $N_{\rm H}$ was fixed to the Galactic value. The
results are summarized in Table 1 along with 90\% errors for single
parameters. Since the data need to be accumulated in a few 100 square
arcmin, the amplitude of temperature variations is suppressed if
small-scale variations exist. The temperatures in the hot and the
medium regions overlap in the 90\% error, but the value in the cool
region (\#2) is significantly lower than the other two. It is notable
that the accumulated spectrum in the east hot region does not show the
high temperature of $\gtrsim 10$ keV seen in the temperature map. This
is because the very hot pixels are almost at the outer edge of the
cluster, where the surface brightness is very low, and the spectral
fit is affected by colder inner pixels. In summary, the spectral fit
confirms the significance of the lower temperature in the extended
region in the east of the cluster center. All the 3 regions indicate a
consistent metal abundance around 0.3 solar.

\section{Discussion}

The mapping observation of the Perseus cluster from ASCA has detected a
remarkable temperature structure as shown in Figure\ 1. Previous
studies from ASCA (Arnaud et al.\ 1994; Ezawa et al.\ 2001) based on
spatially accumulated spectra indicated that the ICM temperature
significantly varied in several regions. However, this is the first
case that the temperature map of the whole cluster is derived with a
spatial resolution $\lesssim 10'$. The Perseus cluster has several
extended cool ($kT \sim 5$ keV) regions which seem to be surrounded by
a filamentary structure with medium ($\sim 7$ keV) to hot ($\sim 10$
keV) temperatures.  This feature looks quite different from, e.g., the
Centaurus cluster, in which a hot region exists near the
cluster center (Churazov et al.\ 1999; Furusho et al.\ 2001).

The remarkable finding is the extended cool region with a low
temperature of about 5 keV at $10'-20'$ east from the cluster
center. As seen in figure 1, this region has a diameter of about $20'$
(600 kpc). The same region was recognized in the ROSAT observations
(Schwartz et al.\ 1992; Ettori et al.\ 1998) based on the substructure
in the surface brightness. The residual image after subtraction of a
$\beta$-model (contours in Fig.\ 1; also Schwarz et al.\ 1992) is very
similar to the structure of the extended cool region. Ettori et al.\
(1998) discuss that there is possibly a group of galaxies merging into
the main body of the cluster. Considering the size and temperature of
the cool region, the merged body may well be an established poor
cluster. No significant variation of metal abundance in this cool
region suggests that the merged cluster was already enriched with
metals to around 0.3 solar. Since the density contrast should be
smoothed out in the sound crossing time ($< 1 \times 10^9$ yr), this
extended cool region would have been created within the recent $10^9$
yr.

Figure 1 also shows that a chain of hot regions seem to be surrounding
the cool region discussed above. It could be that the collision of the
poor cluster, whose direction may be nearly in parallel to the line of
sight, has caused shock heatings and created a large ring-like hot
region. The diameter of the ``hot ring'' is as large as about
$1^\circ$ ($\sim 2$ Mpc). Assuming that the hot region has a
spherical shell structure with inner and outer radii of 500 kpc and 1
Mpc, respectively, its excess thermal energy is estimated to be $2
\times 10^{62}$ erg. The necessary mass of the colliding system is
then at least $2 \times 10^{13} \ M_{\odot}$ if the relative velocity
is about 1000 km sec$^{-1}$. This suggests that a major merger with a
small size cluster (such as Abell 1060) may have created the ring-like
temperature structure.

Let us look into relevant time scales here. The thickness of the ``hot
ring'' is less than about 500 kpc, therefore if the heat front
propagates at $\sim 1000$ km s$^{-1}$ the time required to heat up the
ring is less than $5\times 10^8$ yr, which is comparable to the
ion-electron relaxation time (Takizawa 1999). On the other hand, the
conduction time for this ``hot ring'' structure to be smoothed out is
roughly estimated as $2 \times 10^8$ yr, assuming a standard thermal
conductivity. This means that the hot structure would dissipate away
before the whole region is sufficiently heated up. The presence of the
``hot ring'' suggests that the thermal conductivity in ICM may be
lower by an order of magnitude than the standard value, as recently
pointed out by Ettori and Fabian (2000). In such an outer region of
the cluster with low gas density, however, suppression of the
conduction in terms of strong magnetic field may not be effective.

Other structures in the temperature map are smaller and seemingly
consisting of many blobs with a rough size of $10'$ (300 kpc). We note
that in the Virgo cluster, Shibata et al.\ (2001) reported that the
spatial scale of hot and cool regions are about 300 kpc, which is more
or less the typical size of groups of galaxies. The Perseus cluster
shows hot regions with $kT\gtrsim 8$ keV toward the southern and
northeastern edges of the cluster. Low temperatures ($kT\sim 5$ keV)
are seen at northwestern and southwestern edges, with the latter
around IC 310. These asymmetric temperature structures do not
accompany significant surface-brightness contrast except for the IC
310 subcluster. This indicates that the pressure distribution is not
smooth, and a considerable bulk flow of the gas may be
present. Numerical simulations indicate that the temperature structure
remains several $\times 10^9$ yr after a subcluster collision (e.g.\
Schindler and M\"uller 1993), and the slow relaxation process between
ions and electrons would also cause complex distribution of the
electron temperature (Takizawa 1999).  The observed temperature
structure in the Perseus cluster suggests that this cluster has
experienced many subcluster mergers.  These mergers may well have been
the main building processes which have made the Perseus cluster into
such a massive system, as suggested from numerical simulations of the
cluster growth (Navarro, Frenk, and White 1995).

Finally, the distribution of the cool gas in the southwest near IC 310
looks as if the subcluster gas is blown away from the center of the
Perseus. This suggests that the IC 310 subcluster is now falling into
the main cluster, and the cool gas associated with the subcluster may
be slowed down due to the ram pressure. We hope that observation with
high angular resolution may lead to direct detection of shock
features, such as sharp edges (Markevitch et al.\ 2000) observed with
Chandra, and bring us a closed-up view of the dynamical processes in
the cluster system.

\acknowledgments

We would like to thank Dr. Y. Fukazawa, Y. Ikebe, K. Masai, S. Sasaki,
and Y. Tawara for their support and useful discussion. We also thank 
the anonymous referee for helpful comments and suggestions. T. F. is
supported by the Japan Society for the Promotion of Science (JSPS)
Postdoctoral Fellowships for Research Abroad. This work was partly
supported by the Grants-in Aid for Scientific Research No.\ 12304009
and No.\ 12440067 from JSPS.


\begin{table}
\begin{center}
\caption{Best-fit parameters of the spectral fit for the 3 selected
regions}
\begin{tabular}{lccc} \hline \hline 
&\multicolumn{3}{c}{1 MEKAL model ($N_{\rm H}$ fixed, 2--10 keV)} \\\hline
                 & $kT$  & $Z$    & $\chi^2$/d.o.f. \\ 
region           & [keV] & ~[solar]~ &  \\ \hline
\#1 (west hot) & $7.78^{+0.36}_{-0.30}$ & $0.26^{+0.04}_{-0.04}$ & 90.1/68\\
\#2 (east cool)& $5.77^{+0.19}_{-0.17}$ & $0.34^{+0.04}_{-0.04}$ & 88.8/68\\ 
\#3 (east hot) & $8.27^{+2.01}_{-1.19}$ & $0.38^{+0.18}_{-0.19}$ & 39.7/32\\ 
\hline
\end{tabular}
\end{center}
\end{table}

\begin{figure}[ht]
\begin{center}
\plotone{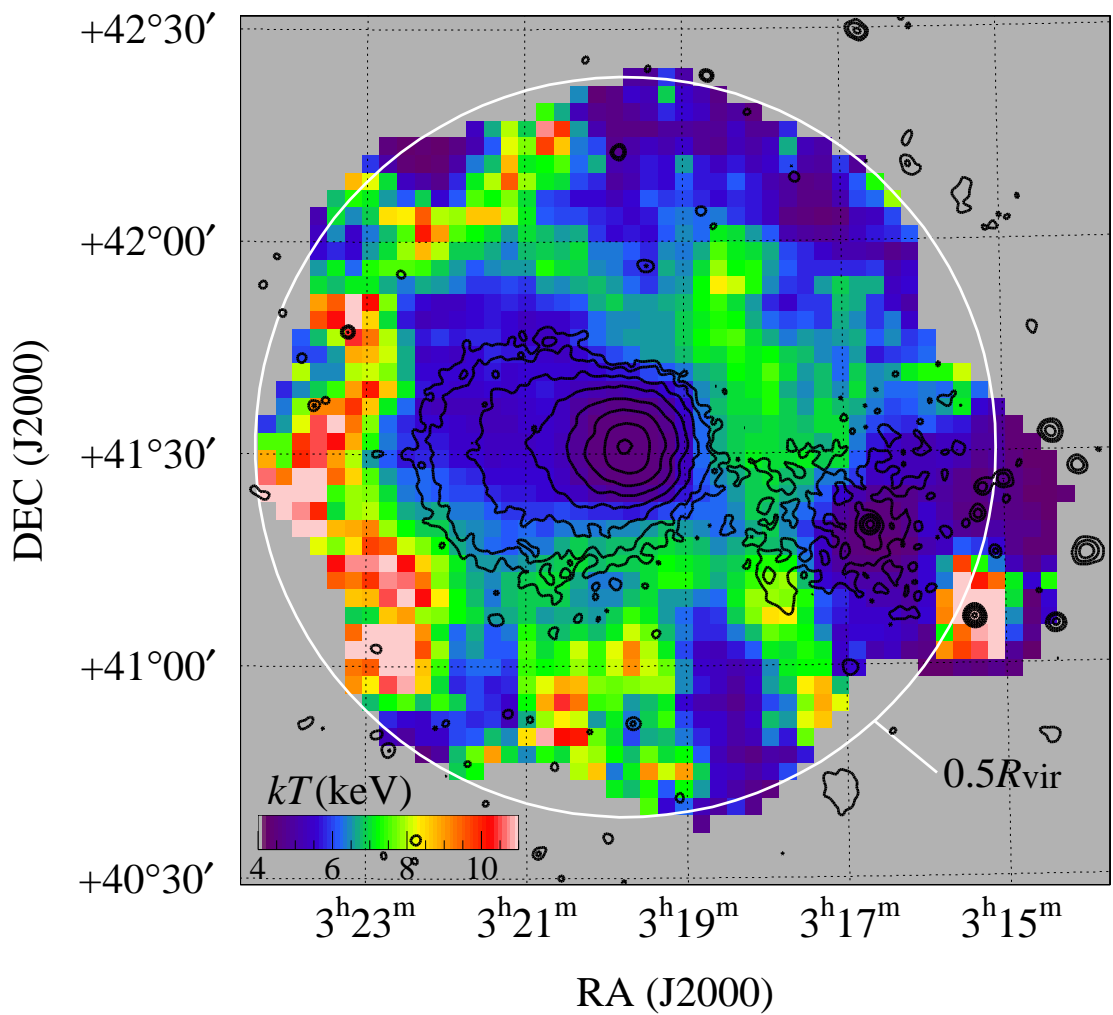}
\caption{ Color-coded temperature map of the Perseus cluster obtained
from ASCA multi-pointing observations. The contours indicate the
residual PSPC image after subtraction of a $\beta$ model. and $R_{\rm
vir}$ is calculated to be $3.3 h_{50}^{-1}$ Mpc for $kT=7$ keV,
assuming the virial radius given as $R_{180}=1.23 (T_{\rm X}/1 {\rm
keV})^{1/2}\ h^{-1}_{50} {\rm Mpc}$ by Evrard, Metzler \& Navarro
(1996). }
\end{center}
\end{figure}

\begin{figure}[ht]
\begin{center}
\plottwo{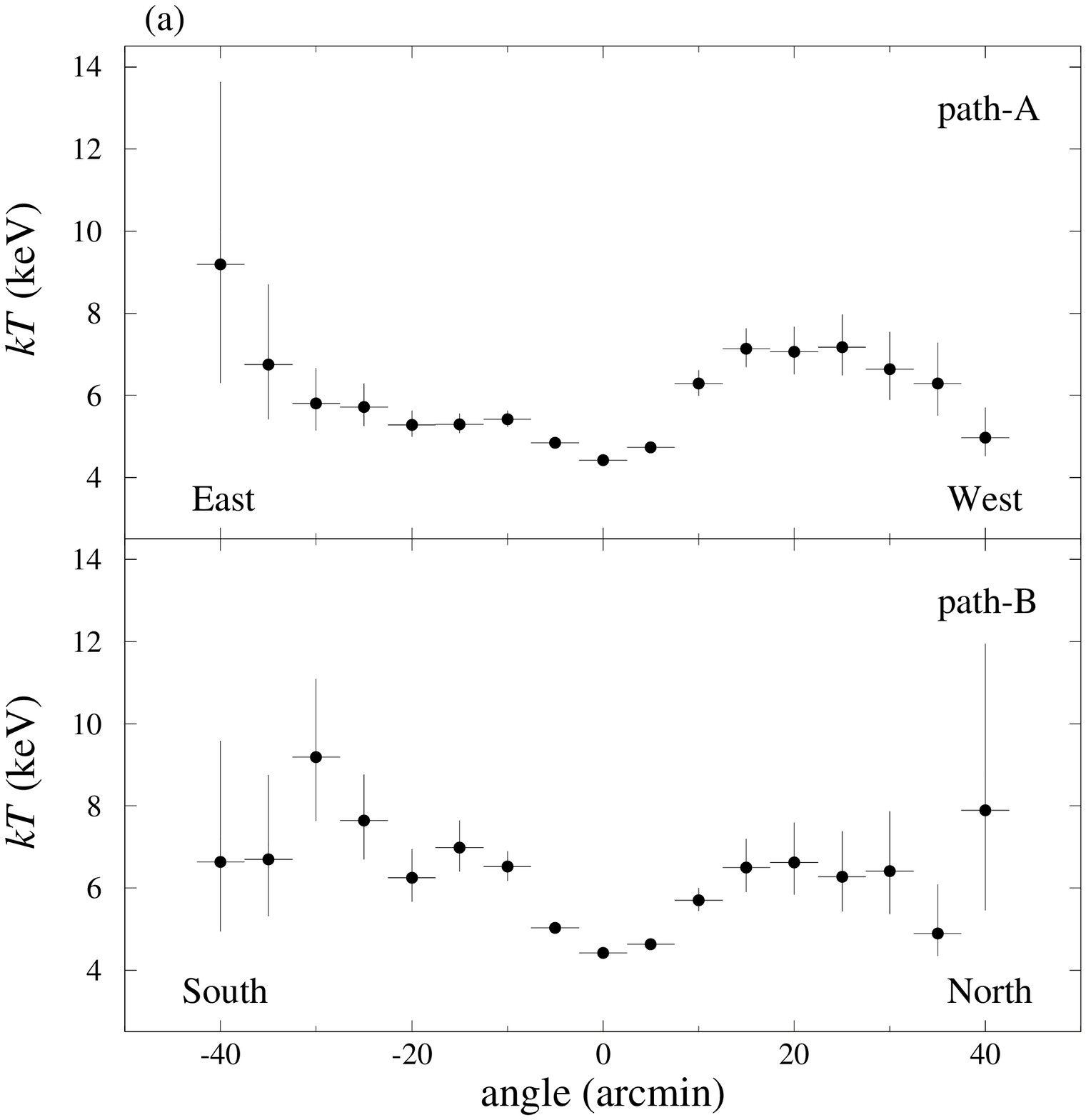}{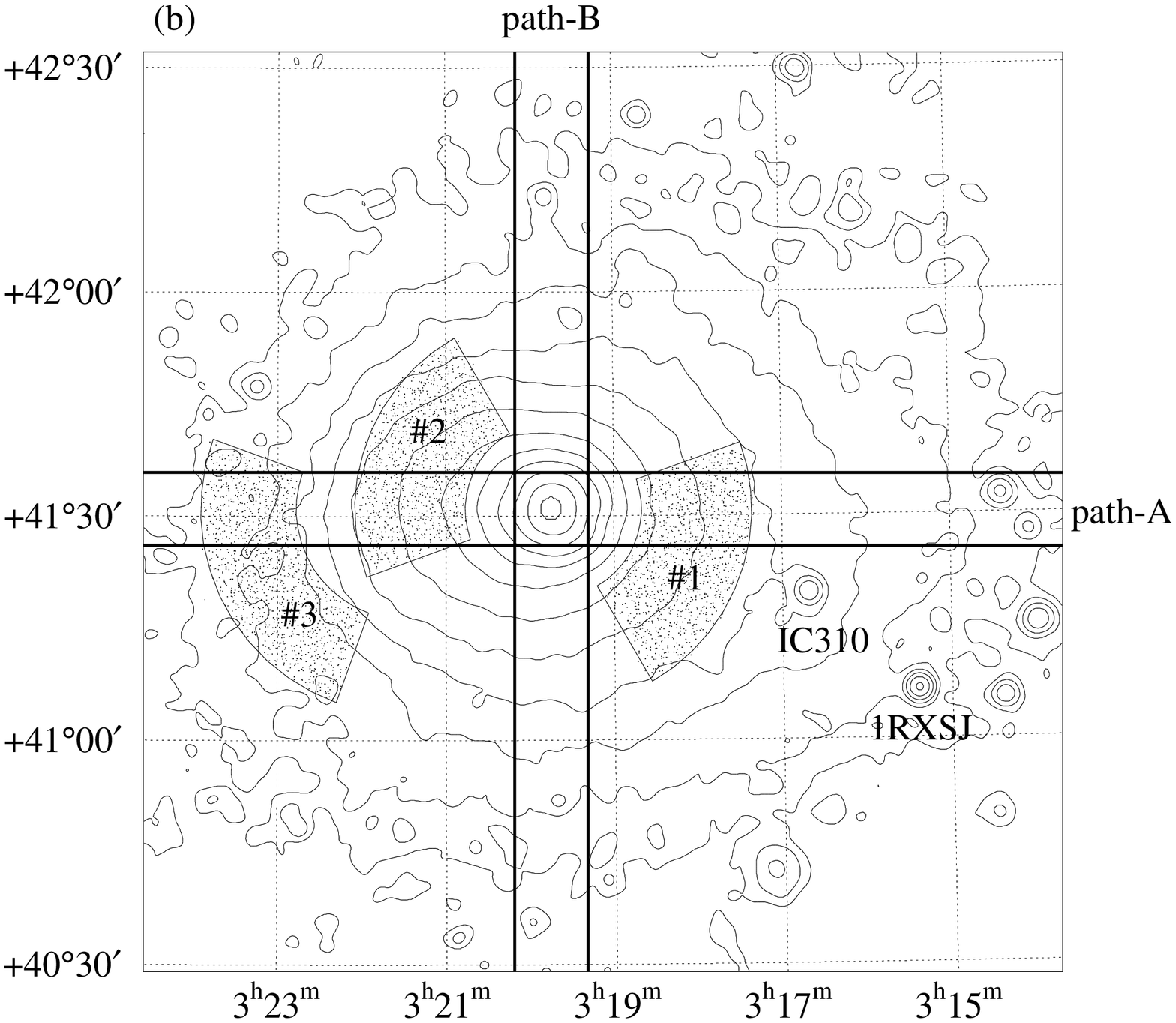}
\caption{Panel (a) shows one-dimensional profiles of temperature along
  the east-west path (A) and north-south path (B), whose positions are
  shown in (b) overlayed on the PSPC contour map. The temperature in
  each bin is calculated for a $10'\times5'$ rectangle with the error
  at the 90\% confidence level. The shaded regions, \#1, \#2 and \#3
  in panel (b), are examined by the spectral analysis, with the
  results shown in Table 1.}
\end{center}
\end{figure}


\begin{references}

\reference{Ander89} Anders, E., \& Grevesse, N. 1989,
Geochim.\ Cosmochim.\ Acta, 53, 197

\reference{Arnau94} Arnaud, K. A. et al.\ 1994, ApJ, 436, L67

\reference{Chu99} Churazov, E., Gilfanov, M., Forman, W., \& Jones, C. 
1999, ApJ, 520, 105

\reference{Ettor98} Ettori, S., Fabian, A. C., \& White, D. A. 1998,
\mnras, 300, 837

\reference{Ettor00} Ettori, S. \& Fabian, A. C. 2000, MNRAS, 317, L57

\reference{Evrar96} Evrard, A. E., Metzler, C. A. \& Navarro, J. F. 1996,
ApJ, 469, 494

\reference{Ezawa01} Ezawa, H. et al.\ 2001, PASJ, No.\ 4 in press

\reference{Fabian00} Fabian, A. C. et al.\  2000, MNRAS, 318, L65

\reference{Furus01} Furusho, T., Yamasaki, N. Y., Ohashi, T., Shibata,
  R., Kagei, T., Ishisaki, Y., Kikuchi, K., Ezawa, H., \& Ikebe,
  Y. 2001, PASJ, 53, 421

\reference{Ikebe95} Ikebe, Y. 1995, Ph.\ D. thesis, University of Tokyo

\reference{Kikuc00} Kikuchi, K. et al.\ 2000, ApJ, 531, L95

\reference{Marke96} Markevitch, M. 1996, ApJ, 465, L1

\reference{Maxim98} Markevitch, M., Forman, W. R., Sarazin, C. L., \&
Vikhlinin, A. 1998, ApJ, 503, 77

\reference{Maxim00} Markevitch, M. et al.\ 2000, ApJ, 541, 542

\reference{NFW95} Navarro, J. F., Frenk, C. S., \& White, S. D. M. 1995,
MNRAS, 275, 720

\reference{Schin93} Schindler, S. \& M\"uller, E. 1993, A\&A, 272, 137

\reference{Schwr92} Schwarz, R. A., Edge, A. C., Voges, W.,
B\"ohringer, H., Ebeling, H., \& Briel, U. G. 1992, A\&A, 256, L11

\reference{Shiba01} Shibata, R. et al. 2001, ApJ, 549, 228

\reference{Snowd94} Snowden, S. L., McCammon, D., Burrows, D. N., \& 
Mendenhall, J. A.\ 1994, \apj, 424, 714

\reference{Takiz99} Takizawa, M. 1999, ApJ, 520, 514

\reference{Tsusa95} Tsusaka, Y., et al.\ 1995, Appl.\ Opt., 34, 4848

\reference{Watan99} Watanabe, M., Yamashita, K., Furuzawa, A.,
  Kunieda, H., Tawara, Y., \& Honda, H. 1999, ApJ, 527, 80

\reference{Watan01} Watanabe, M., Yamashita, K., Furuzawa, A.,
Kunieda, H., \& Tawara, Y.\ 2001, PASJ, No.\ 4, in press
\end{references}
\end{document}